\documentclass[pre,aps,twocolumn]{revtex4}
\usepackage{epsfig}
\usepackage{bm} 

\newcommand{\Onecol} {\begin{widetext} \onecolumngrid} %% 2 -> 1
\newcommand{\Twocol} {\end{widetext} \twocolumngrid} %% 1 -> 2
\newcommand{\be}{\begin{equation}}
\newcommand{\ba}{\begin{array}}
\newcommand{\bea}{\begin{eqnarray}}
\newcommand{\bfi}{\begin{figure}}
\newcommand{\ee}{\end{equation}}
\newcommand{\ea}{\end{array}}
\newcommand{\eea}{\end{eqnarray}}
\newcommand{\efi}{\end{figure}}

\begin{document} 
\bibliographystyle{prsty}
\title{Flow reversal in a simple dynamical model of turbulence}
\author{Roberto Benzi}
\affiliation{Dip. di Fisica and INFN, Universit\`a ``Tor
Vergata", Via della Ricerca Scientifica 1, I-00133 Roma, Italy.} 
%\pacs{47.27-i, 47.27.Nz, 47.27.Ak}
\begin{abstract} 

In this paper, we study a simple hydrodynamical model showing abrupt flow reversals at random times. For a suitable range of
parameters, we show that the dynamics of flow reversal is accurately described by stochastic differential equations, where the noise
represents the effect of turbulence. 

\vskip 0.2cm 
\end{abstract} 
\maketitle 
%%%%%%%%%%%%%%%%%%%%%%%%%%%%%%%%%%%%%%%%%%%%%%%%%%%%%%%%%%%%%%%%%%%%%%%%%%%%

It has been recently reported \cite{sreeni} that abrupt flow reversal takes place at large Rayleigh
number in thermal convection. 
Beside thermal convection, flow reversal has been also observed in laboratory
experiment of two dimensional turbulence \cite{sommeria} and in the magnetic polarity of the earth \cite{mhd}.
More
generally, there are many "turbulent" flows for which transitions between
different states have been investigated, namely within the theory of multiple equilibria for 
atmospheric flows  \cite{ghil} and 
of long time climatic changes \cite{sutera}.

In most cases the major question to be answered by experiments or observations concerns the mechanism responsible for the 
transition. This question is highly non trivial whenever the average persistent time $<\tau>$ around different states is much
longer than any characteristic times describing the dynamical behavior of the system.

There are two main interpretations which have been suggested so far. 
The first one assumes that turbulence is more or less a "noise" applied to the order parameter $\psi$ 
which describes the system (i.e. the wind in the case of RB convection or the temperature in the case of climatic change). 
Then, by defining $\psi$ such that the two observed states are $\pm \psi_0$,
the equation for $\psi$ is given by:
\begin{equation}
	d\psi = [a\psi(1 -\frac{\psi^2}{\psi_0^2})]dt + \sqrt{\sigma} dW(t)
\label{dphi}
\end{equation}
where $W$ is a Wiener process $\delta$ correlated in time while $a^{-1}$ is the characteristic time scale of the instability
at $\psi = 0$.
One can show that transitions between 
the two stable states $\pm \psi_0$ occur at random times $\tau$  with an average time $<\tau>$ given by
\begin{equation}
	<\tau> \sim \frac{\pi}{\sqrt{2}a}exp(a\psi_0^2/(2\sigma))
\label{tau}
\end{equation}
Hereafter, following the 
language of stochastic differential equations, the random time $\tau$ will be referred to as ``exit time''.
Note that equation (\ref{dphi})
implies, for small $\sigma$, that the average exit time $<\tau>$ is much longer than the deterministic
"fluidodynamical" time $a^{-1}$. Moreover, by employing the theory of stochastic differential equations \cite{gs}, one can
compute  the probability
distribution of the exit time $\tau$, which, for small $\sigma$, is given by:
\begin{equation}
	P(\tau) = <\tau>^{-1} exp(-\tau/<\tau>)
	\label{pdftau}
\end{equation}
If the above scenario is believed to be correct, 
then the transitions between the two states $\pm \psi_0$ are due to 
repeated small noise perturbations of the same "sign" which are acting against the deterministic "force", i.e. there is
no specific mechanism introduced by 
the small scale turbulence (parametrized by the noise) and transitions can be
explained in terms of large deviation theory.

One of the major criticism against the above interpretation is that 
the noise \textit{is} by itself the effect of turbulence and, in most cases, 
there is no time scale separation between the dynamic fluctuations of $\psi$ 
and turbulent fluctuations. Thus one cannot assume that ``the noise'' is a ``fast'' perturbutaion
with respect to the dynamics of $\psi$ and, as a consequence,  equation (\ref{dphi}) cannot be   
justified. As an alternative way, one should look for a 
specific "fluidodynamical" large scale mechanism which can explain the observed transitions.
For instance, for the wind reversal in thermal convection, there has been
a recent proposal \cite{detlef} which explains transitions as the result of plume dynamics.

In this letter we want to understand  
whether and how equation (\ref{dphi}) can be justified, at least 
in  the simplest possible model of a  "turbulent" flow. 
For this purpose we shall consider an  'energy cascade' model i.e. 
a shell model aimed at reproducing few of the relevant characteristic features of the  
statistical properties of the Navier Stokes equations \cite{fri95},\cite{bohr}.
In a shell models, the basic variables describing the 'velocity field' at scale 
$r_n = 2^{-n} r_0 \equiv k_n^{-1}$, is a complex number $u_n$ satisfying a suitable set of non linear equations. There are many 
version of shell models which have been introduced in literature. Here we choose the one referred to as {\em Sabra} shell model
\begin{eqnarray}
&&\frac{d u_n}{dt} =   i k_n [a \Lambda u_{n+1}^*u_{n+2} 
+ b u_{n-1}^*u_{n+1} -c \Lambda^{-1} u_{n-2}u_{n-1} ] \nonumber \\
&& - \nu k_n^2 u_n + f_n  
  \label{sabra}
\end{eqnarray}
where $\Lambda=2$, $a=1$ and $c= -(1+b)$ and $f_n$ is an external forcing. 
Let us remark that the statistical properties of intermittent fluctuations, computed
either using $u_n$ or the instantaneous rate of energy dissipation, are in close {\em qualitative} and {\em quantitative} agreement
with those measured in laboratory experiments, for homogeneous and isotropic turbulence \cite{bif03}. 

The basic idea of our approach is to assume that $U_r \equiv real(u_1)$ describes the one dimensional unstable manifold arising by 
a (large scale) pitchfork bifurcation. Consequently
we change the equation for $u_1$ as follows:
\begin{eqnarray}
&&\frac{d u_1}{dt} =   \Phi +  \mu u_1(1-\frac{u_1^2}{u_0^2})  - \nu k_1^2 u_1   
  \label{sabrau1} \\ 
&&\frac{d u_n}{dt} =   i k_n [a \Lambda u_{n+1}^*u_{n+2} 
+ b u_{n-1}^*u_{n+1} -c \Lambda^{-1} u_{n-2}u_{n-1} ] \nonumber \\
&& - \nu k_n^2 u_n + f_n  \,.\,\,\,  (n>1)
  \label{sabraun}
\end{eqnarray}
where $\Phi \equiv ik_1 a \Lambda u_{2}u_{3}^*$.
Let us comment equation (\ref{sabrau1}). 
In most cases, a pitchfork bifurcation, 
as described by equation (\ref{sabrau1}), is observed in real fluidodyanamical flows with respect to the external forcing or
the Reynolds number. In our case we are assuming that the unstable manifold is
 coupled to smaller scales by 
the term $ i k_1 a \Lambda u_{2}^*u_{3}$.  
For small $\nu$, the two states $u_1 = \pm u_0$ 
becomes unstable and a turbulent regime is observed. In the following  we will think of equation 
(\ref{sabrau1}) as a realistic, although approximate, 
equation describing a simplified turbulent "flow" superimposed to a large scale instability. 
As one can see, no \textsl{external noise} is introduced in the system. 

We remark that equations (\ref{sabrau1}-\ref{sabraun}) are symmetric under the transformation $u_1 \rightarrow -u_1$. More precisely, one can show that by 
changing $(u_{3m+1},u_{3m+2},u_{3m+3})$ with $(-u_{3m+1},-u_{3m+2},u_{3m+3})$ ( $m=0,1,2,..$), the equations of motion
are invariant.
Using dimensionless variables $W = u_n/u_0$, $K_n = k_n L$ and $t' = \mu t$ ($L\equiv k_1^{-1}$), one gets:
\begin{equation}		
\frac{d W_1}{dt'} =   i  \frac{u_0}{\mu L}K_1 a \Lambda W_{n+1}^*W_{n+2} +  W_1(1-W_1^2)  - \frac{\nu}{\mu L^2}  K_1^2 W_1   
  \label{sabraw}
\end{equation}
Equation (\ref{sabraw}) tells us 
that the dynamical behavior of $u_n$ depends on two dimensionless number, 
namely the Reynolds number $Re \equiv u_0L/\nu$ and the number $B \equiv u_0/(\mu L)$. 
We will investigate the statistical properties of eq. (\ref{sabrau1}-\ref{sabraun}) for 
$Re \rightarrow \infty$ and for different values of $B$. In particular we fix $\mu=1$ and $u_0$ real, while the parameters of the model
are  $a=1,b=-0.4, c=0.6$. 

\begin{figure}[h]
	\begin{center}
		\includegraphics[width=0.50\textwidth]{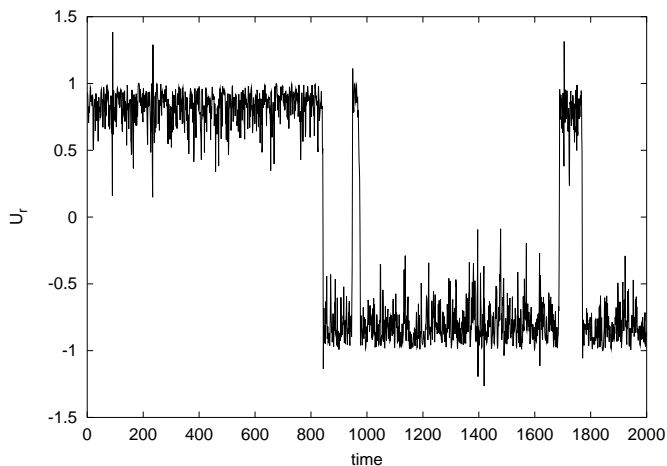}
	\end{center}
	\caption{The velocity $U_r$ plotted as a function of time as obtained by numerical simulation of equation(\ref{sabrau1}) for $u_0=1$, $\mu=1$ and $\nu = 10^{-6}$.}
	\label{fig1}
\end{figure}

\begin{figure}[h]
	\begin{center}
		\includegraphics[width=0.50\textwidth]{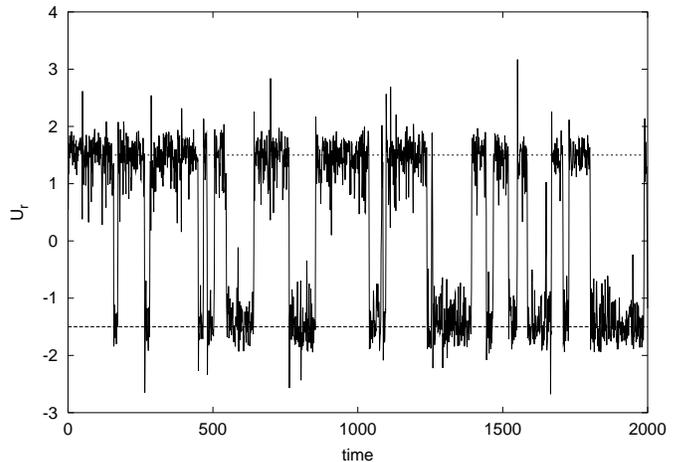}
	\end{center}
	\caption{Same as in figure (\ref{fig1}) for $u_0=2$.}
	\label{fig2}
\end{figure}

\begin{figure}[h]
	\begin{center}
		\includegraphics[width=0.50\textwidth]{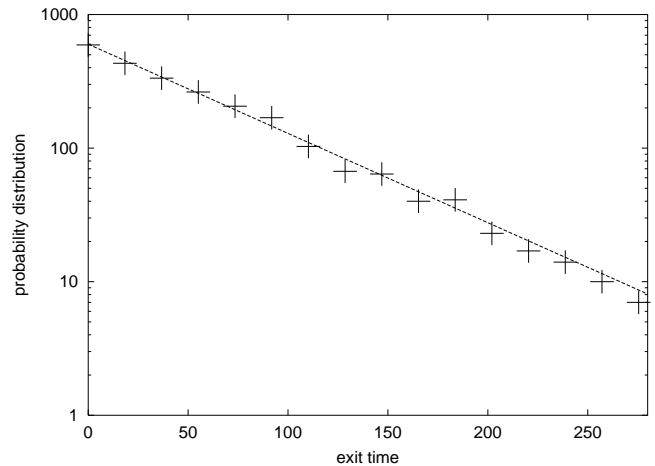}
	\end{center}
	\caption{$log[P(\tau)]$ as a function of $\tau$ as obtained by numerical simulations of (\ref{sabrau1}) for $u_0=1$. According
to the theory of stochastic differential equations, $P(\tau)$ is $exp(-\tau/<\tau>)$ where $<\tau>$ is the average exit times. The line
in the figure shows  $exp(-\tau/65)$ which is an extremely good fit to the observed $P(\tau)$.}
	\label{fig3}
\end{figure}

In figure (\ref{fig1}) 
we show $U_r$ 
as a function of time for $\nu=10^{-6}$ and $B=1$ ($u_0=1$), while in figure (\ref{fig2}) we show the
same variable for $B=2$ ($u_0=2$).  
As one can see abrupt reversals of $U_r$ are observed at apparently random times in both cases.     
The most important feature of figure (\ref{fig1}) and (\ref{fig2}) 
is that the characteristic correlation time of $U_r$  
is of order $1$ (i.e. it is of order $L/u_0 \sim 1$), much smaller than the exit time $\tau$. 
The behavior shown in figures (\ref{fig1}-\ref{fig2}) does not change by increasing the {\em Reynolds} 
number.
A more refined numerical simulation shows that 
the "random" exit times $\tau$ are distribute according to eq.(\ref{pdftau}) 
with $<\tau> \sim 600$ and $<\tau> \sim 65$ for $B=1$ and $B=2$ respectively. 
In figure (\ref{fig3}) we show $ log P(\tau)$ versus 
$\tau$ for the case $B=2$, where the line shown in the figure represent the quantity $exp(-\tau/65)$.

For larger values of $u_0$, the average exit time $<\tau>$ becomes smaller and eventually, for $u_0 \sim 10$ it becomes
of order $1$ (see also the discussion below). 
Our interest will focus for values of $u_0$ in the range $[1,2]$ where $<\tau>$ is at least two order of magnitude 
longer than $\mu^{-1}$ and $L/u_0$, the two relevant deterministic time scales of equation (\ref{sabrau1}).

A closer look at figures (\ref{fig1}) and (\ref{fig2}) reveals that the two states of $U_r$ are not $\pm u_0$. For $B=1$ the maxima 
$\pm U_M$
in
the probability distribution of $U_r$ are located at $U_M = 0.84$, while for $B=2$ we find $U_M = 1.5$, represented as horizontal
 lines in
figure (\ref{fig2}).

In order to explain these results, let us consider more carefully the physical 
meaning of $\Phi$ in equation (\ref{sabrau1}). The quantity
$real(\Phi u_1^*)$ is the amount of energy transferred by mode $u_1$ to smaller scales, i.e. $u_2$ and $u_3$. 
On the average, we know that $real<(\Phi u_1^*)> \equiv - \epsilon < 0$, where $\epsilon$
is the average rate of energy dissipation. Thus, as a first approximation, we can assume that
\begin{equation}
	\Phi = - \beta u_1 + \phi'
	\label{phi}
\end{equation}
where $<\phi'u_1^*>=0$ and  $\beta >0$.    
By multiplying both side by $u_1^*$, taking the time average, we can compute $\beta$ as: 
\begin{equation}
	\beta = real(<\Phi u_1^*>)/<|u_1|^2>
	\label{calcolabeta}
\end{equation}
Thus we must expect that two maxima in the probability distribution of $U_r$ should corresponds to the solution of the 
equation:
\begin{equation}
(\mu - \frac{\epsilon}{<|u_1|^2>}) U_r - \frac{\mu}{u_0^2}U_r^3 = 0.
\label{fixpoint}
\end{equation}
Equation (\ref{fixpoint}) tells us two important informations. First, the ``states'' $\pm U_M$, 
(between which abrupt transitions are observed) are
not stationary solutions of the deterministic equations (\ref{sabrau1}), rather the ``states'' should be considered as 
{\em statistically stationary states} of the system. Second, we must expect $U_M <u_0$ 
as far as an energy cascade,  $\epsilon > 0$, is produced.
\begin{figure}
	\begin{center}
		\includegraphics[width=0.50\textwidth]{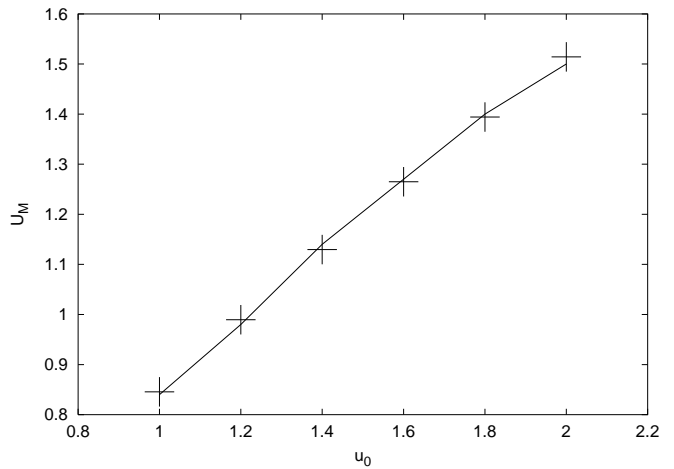}
	\end{center}
	\caption{Computing the statistically stationary solutions for $U_r$. The line corresponds to $u_0 \sqrt{\mu-\beta(u_0)}$, where
$\beta(u_0)$ is defined by equation (\ref{calcolabeta}). The symbols represent the value of the maxima $U_M$ of the probability
density function of $U_r$ obtained by the numerical simulations of (\ref{sabrau1}) for different values of $u_0$.}
	\label{fig4}
\end{figure}
We have carefully checked the validity of equation (\ref{fixpoint})  for  $u_0$ in the range $[1,2]$. We have computed $\beta$ from 
numerical simulations according to equation (\ref{calcolabeta}). It turns out that the numerical values of $\beta$ are extremely
well represented by the function $\beta(u_0) = 0.15+0.14u_0$.
Next we have solved equation (\ref{fixpoint}) for each value of 
$u_0$ and $\beta(u_0)$ getting the line represented in figure (\ref{fig4}). Finally we have computed the value of $U_M$ using  the
the probability 
distribution of $U_r$ as obtained by the numerical simulations of the
model. The results are shown as symbols in figure (\ref{fig4}): the agreement is excellent.

At the light of the above results, it is tempting to argue that the behavior of (\ref{sabrau1}) is consistent with the stochastic
differential equation:
\begin{equation}
 d u_1 =    [(\mu-\beta(u_0)) u_1 + \mu u_1 \frac{u_1^2}{u_0^2}]dt  + \sqrt{\sigma} dW(t) 
\label{sabranoise}
\end{equation}
for a suitable value of the noise variance $\sigma$. Using (\ref{phi}) and (\ref{sabrau1}), the quantity $\sqrt{\sigma}dW(t)$ represents
the {\em deterministic} term $\phi'(t) dt$. In order to validate (\ref{sabranoise}), we need to check whether the observed
fluctuations of $u_1$ are in agreement with the average exit time $<\tau>$. More specifically, using (\ref{sabranoise}) and (\ref{tau})
we have:
\begin{equation}
 <\tau> = \frac{\pi}{\sqrt{2} (\mu-\beta)} exp((\mu-\beta)^2u_0^2/2\mu\sigma)
\label{tausigma}
\end{equation}

By the numerical value of $<\tau>$ and $\beta$, we can compute the amplitude of the noise $\sigma$,
hereafter denoted by $\sigma_{\tau}$, needed
to explain the observed average exit times. The crucial point is 
whether the observed fluctuations of $U_r$ near one the two states are ``compatible'' with the noise amplitude
$\sigma_{\tau}$.  To answer this question,  let $\delta U$ be  
small deviation around the statistically stationary
states. Using (\ref{sabranoise}) we can compute ${}$
$<\delta U)^2>$ as:
\begin{equation}    
<(\delta U)^2> = \frac{\sigma}{4(\mu-\beta)}
\label{sigma1}
\end{equation}  
By using the numerical simulations, 
we can estimated ${}$
$<(\delta U)^2>$  around the statistically stationary states. Finally, 
using (\ref{sigma1}), we can estimated
the ``noise'' variance $\sigma$, hereafter referred to as $\sigma_{\delta U}$, which explains the observed fluctuations of
$\delta U$. 
For equation (\ref{sabranoise}) to represent a good approximation of the full non linear deterministic
system, we must obtained
\begin{equation}
\sigma_{\delta U} \sim \sigma_{\tau}
\label{controllo}
\end{equation} 
\begin{figure}%[h]
	%\begin{center}
\includegraphics[width=0.50\textwidth]{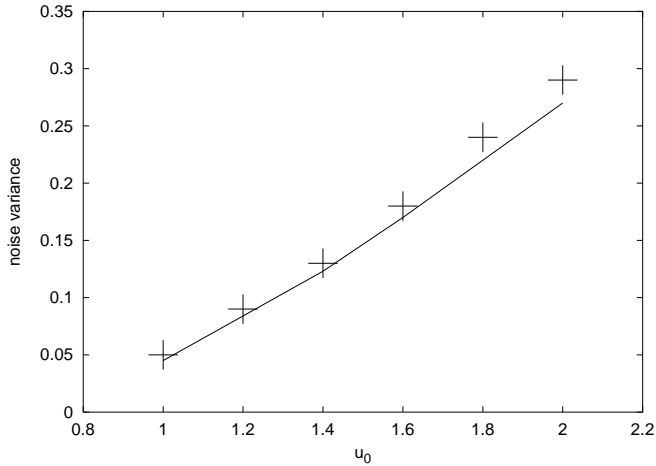}
	%\end{center}
\caption{Plot of $\sigma_{\delta U}$ (symbols) and $\sigma_{\tau}$ (line) for different values of $u_0$.$\sigma_{\tau}$ is the
noise computed by equation (\ref{tausigma}) while $\sigma_{\delta U}$ is the noise computed from the fluctuations of $U_r$ near the
statistically stationary states, see equation (\ref{sigma1}). }
\label{fig5}
\end{figure}
In figure (\ref{fig5}), we plot $\sigma_{\tau}$ (line) and $\sigma_{\delta U}$ (crosses) for different values
of $u_0$ in the range $[1,2]$. As one can see, (\ref{controllo}) is verified
with very good accuracy,  with at most {\em 10 per cent} difference for $u_0=2$.
We remark that the result shown in figure (\ref{fig5}) represent a rather severe test on the validity of equation (\ref{sabranoise}).

We want finally to comment  on the behavior of the model in the limit of large $u_0$ or, equivalently, in the limit of large $B$.
For large value of $u_0$ we reach the condition $\beta(u_0) \sim \mu$. 
Actually, arguments based on energy balance and numerical simulations clearly
show that the maximum value of $\beta$ is $1$. 
In this case, equation (\ref{fixpoint}) gives $U_M \sim 0$, i.e.
the two ``statistically stationary states'' disappear.  
It follows that, for large $B$, one cannot speak of ``abrupt flow reversal''.

All the results, presented so far, strongly indicate that equation (\ref{sabranoise}) is a very good candidate to explain
the observed abrupt ``flow reversal'' shown in figures (\ref{fig1}) and (\ref{fig2}), i.e. ``flow reversal'' is explained within
the  framework of large deviation theory of stochastic differential equations. Therefore, we argue that 
some of the experimental results mentioned in the
introduction, can be investigated and explained
in the framework of stochastic differential equations 
even if no time scale separation exists between the ``wind'' fluctuations and
turbulence.

\acknowledgments \vskip 0.2 cm The author thank 
L. Biferale, F. Toschi, M. Sbragaglia and I. Procaccia for useful discussions.

%\bibliography{/home/biferale/PAPERS/PhysRep/Style/literatur}

\end{document}